# Early Evolution of Bird-Type Language without Grammar: Duplication and Mutation

Miki Fukunoue and Osamu Narikiyo*

*Department of Physics, Kyushu University, Fukuoka 819-0395, Japan*

ABSTRACT

Using a series of computer simulations we have demonstrated a scenario of the early evolution of the bird-type primitive language. We do not assume wise agents who can use a grammar and manage an evolution without a grammar. Duplication and mutation of phrases is our strategy. Such a strategy is seen in wide classes of living phenomena.

* Corresponding author.
 *E-mail address:* narikiyo@phys.kyushu-u.ac.jp (O. Narikiyo)

## 1. Introduction

Our language of human beings is analyzed by a grammar (Chomsky 2002). On the other hand, the first language on our planet is supposed to be established without a grammar, since the users of the language are not so wise. Thus we present a scenario of an early language evolution of a primitive language without a grammar. It should be noted that we see a grammar whenever a language is used even if the users do not utilize a grammar.

The origin of a language is thought to be an establishment of an index-correspondence between events and signals (Deacon 1997; See especially Fig. 3.3.). The mechanism of such an establishment is not explained at present (we can only describe the situation from outside: Izutsu 2011) so that we assume it and only discuss the evolution after it.

Our target is a bird-type primitive language. Then in the next section we review observations and hypotheses of a bird language.

## 2. Bird Language

Observations show that birds have languages. Some language-learning processes through generations are reported (Sasahara et al. 2007):
- A child starts the learning by mimicking the utterances of the others.
- The patterns of the utterance are gathered from several masters.
- A pattern is handled as a group of phrases. The order or combination of the phrases is edited by the learner.

There are two hypotheses to explain above learning processes (Sasahara et al. 2007). One is the word-first hypothesis where sentences are composed from words under the circumstance where each word has its own meaning. The other is the sentence-first hypothesis where the correspondence between a story and a sentence is established firstly. Then through comparison of sentences a meaning is assigned to a common phrase. Such a phrase is a word.

In either hypothesis some wise agents are needed to accomplish the above tasks. In the former agents can form and relate categories. In the latter agents can extract words. On the other hand, we are interested in the most primitive language even agents lacking such abilities can use. Then we perform a series of computer simulations without wise agents.

On the one hand, now-existing birds have become too wise through long-term evolution to neglect some ability to use grammar. On the other hand, our agents before a grammar are expected to correspond to some creature which once used the almost first language on our planet and became extinct.

## 3. Duplication and Mutation

We stand for the sentence-first hypothesis. However, the present proposition for the bird language is based on the observation of now-existing birds so that it has far evolved from the first language. Then we pursue the most primitive scheme for the imaginary first language.

Our strategy is duplication and mutation. Such a strategy works at various hierarchies of living systems. At the bottom the evolution of the gene is driven by this mechanism (Ohno 1970). At the middle this is the basic body plan in developmental biology (Hall 1992). At the top we employ the same for the evolution of the first language.

## 4. Model Simulations

Basic idea of our model is duplication and mutation. We implement such an idea in the simplest manner.

The following two are the preconditions for our model:
- Several patterns of repeated phrases have their own meanings and work as sentences.
- Phrases are determined by physical conditions of utterance without meaning.

Using the series of simulations we discuss the following issues:
a. A pattern of the repetition of a phrase changes into a mixture of phrases by mutation. This gives a ground for the generation of words from sentences.
b. Language is defined by the society.
c. Diversity arises from shuffling of order. (The diversity of genes is also generated by exon shuffling.)
d. Category formation is not necessary. The degeneracy of the many to one correspondence (several meanings are represented by one signal due to the poverty of vocabulary) works as a pre-category.

## 4a. Error in Parent-Child Learning

As the smallest environment of the language learning we model a learning activity between a parent and a child. The development of the learning here is driven by the minimization of the cost of the utterance.

First we summarize our scenario for the parent-child learning:
- The learning is performed with some error. A sentence acquired by the child can be deformed from that of the parent unless the deformation has enormous influence on the communication between them.
- The parent has a tolerance for the variant, the deformed language of the child. If the cost of the utterance of the variant is smaller than that of the original, the variant has a larger chance to be used than the original.
- Under the situation where both the original and the variant are used, the possibility arises that these two sentences are linked to similar but slightly different meanings. We regard this branching as the mechanism to produce words from sentences.

Our simulation for the parent-child learning is as follows.
(1) We consider the learning between a parent and a child. We start with a set of a parent and a child at the first generation. The child at the first generation becomes the parent at the second generation who has only one child. The child at the second generation becomes the parent at the third generation who has only one child. We continue this one-dimensional sequence over 1000 generations.
(2) The parent at the first generation teaches three kinds of sentences to the child. These original sentences are (0,0), (1,1) and (2,2) each of which is the shortest representation of a repetition of a phrase. We assume that a repetition of a phrase works as a sentence. For example, the sentence (1,1) is constituted of the repetition of the phrase 1. The meanings of the sentences, (0,0), (1,1) and (2,2), are labelled by X, Y, and Z, respectively.
(3) We assume that a child understands the meaning of a sentence by its first phrase and is not sensitive to its second phrase. Consequently for a child not only (0,0) but also (0,1) and (0,2) have the meaning X. Similarly not only (1,1) but also (1,0) and (1,2) have the meaning Y and not only (2,2) but also (2,0) and (2,1) have the meaning Z.
(4) We assume that the costs of the utterance for 0, 1 and 2 differ. The costs are 0, 1 and 2 for the phrases, 0, 1 and 2, respectively.
(5) The child at each generation acquires three sentences, (0,*), (1,*) and (2,*) where *= 0 or 1 or 2, which have the meaning X, Y and Z, respectively. At the beginning of the learning the child hears the utterance of the parent and mimics it by randomly choosing the second phrase of each sentence. The child presents the chosen sentence to the parent. If the parent accepts it, the learning of the sentence is accomplished and the sentence is fixed. The child teaches the fixed sentence at the next generation as the parent.

(6) The tolerance of a parent is set as follows. Unless its cost exceeds that of the utterance of the parent, the parent accepts the chosen sentence by the child. Otherwise the parent accepts the chosen sentence by the probability $P[\Delta E] = \exp(-\Delta E / T)$ where $\Delta E$ is the increase of the cost and $T$ is the degree of the tolerance. For example, if the utterance of the parent is (1,1) and the chosen sentence by the child is (1,2), $\Delta E = 2 - 1 = 1$.

(7) We have repeated the above procedure up to 1000-th generation.

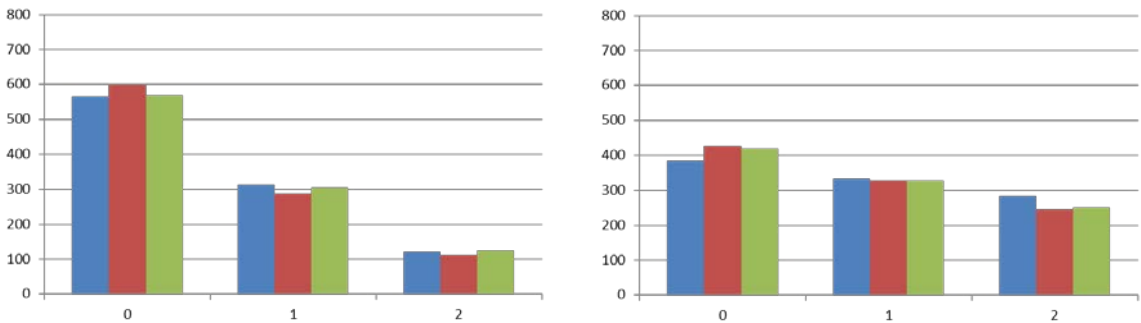

**Fig. 1** Histograms of accepted sentence. (Left) $T = 1.44$ where the variant with $\Delta E = 1$ is accepted by about 50% probability and $\Delta E = 2$ variant by about 25%. (Right) $T = 4.48$ where the variant with $\Delta E = 1$ is accepted by about 80% probability and $\Delta E = 2$ variant by about 64%. Horizontal axis represents the second phrase of each sentence. Vertical axis represents the frequency of the acceptance over 1000 generations. Blue, Red and Green bars represent the frequencies for the meanings X, Y and Z, respectively.

The results of the simulation are shown in Fig. 1. We see that the variants are easily accepted in parent-child learning.

The learning in this section is driven by the minimization of the cost of the utterance for simplicity. Of course the other factors can drive the learning. For example, the maximization of the pleasure of the utterance can be an important factor of the learning.

## 4b. Selection by Society

As seen in 4a the variants are easily accepted in the smallest learning environment. However, a language is a common tool of a society, a larger environment. Thus the variants should be selected by the society.

First we summarize our scenario for the selection:
- A child acquired domestic language takes part in the communication in the society.
- There exists a selection pressure by the society where the use of majority's language is advantageous in the communication.
- However, if the gap between domestic and majority's languages is large, it becomes hard for the child to master the majority's language.
- Every child teaches the language learned in the society at the next generation as the parent.

Our simulation for the selection by the society is as follows.
(1) The learning between a parent and a child is the same as that in 4a. We consider $N$ families in parallel.
(2) After this domestic learning all $N$ children experience the learning in the society where every child performs $C$ times feedback as follows.
(3) In one feedback process every child
   A) check the frequencies, $F_0$, $F_1$ and $F_2$, of the second phrase of each sentence where $F_i$ ($i = 0,1,2$) represents the number of the other children who use $i$ in the second phrase now
   B) and select the second phrase of each sentence by the probabilities, $P_0$, $P_1$ and $P_2$, where $P_i = F_i /(N-1)$.
   C) Each selection is accepted by the probability $P[\Delta D] = \exp(-\Delta D / T_{soc})$ where $\Delta D$ is the difference between the selected phrase $i$ and the domestically learned phrase $j$ ($j = 0,1,2$) defined by $\Delta D = |i - j|$.
   D) The processes B and C are repeated until all second phrases of three sentences are accepted.
(4) After the selection process (3) for the child $C_n$ the frequencies, $F_0$, $F_1$ and $F_2$, for the next child $C_{n+1}$ changes from those for $C_n$ in general.
(5) After the experience of the learning in the society every child becomes the parent of the next generation and starts the teaching process to the child as (1).
(6) We have repeated the above procedure up to 1000-th generation.

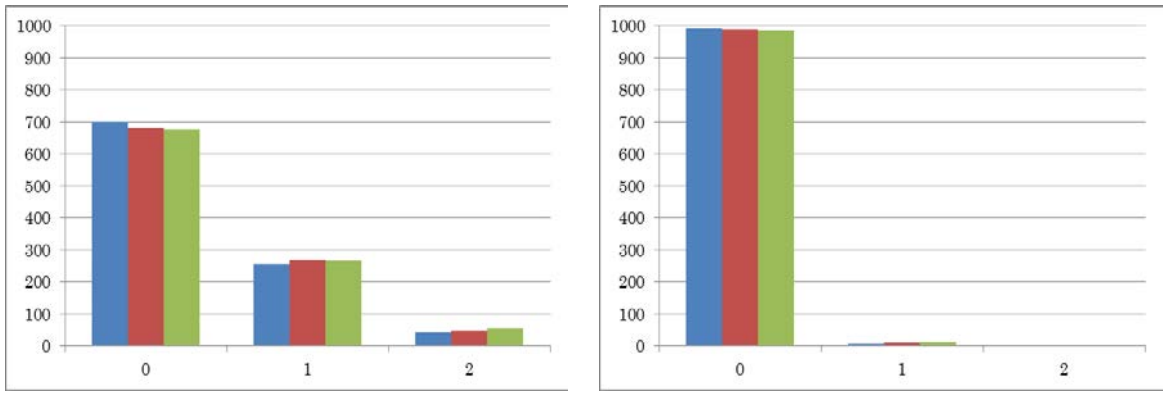

**Fig. 2** Histograms of accepted sentence for $C = 5$ with (Left) $N = 10$ and (Right) $N = 100$. The society ($T_{soc} = 0.83$) is less tolerant than families ($T = 1.44$). Domestically the variant with $\Delta E = 1$ is accepted by about 50% probability and $\Delta E = 2$ variant by about 25%. Socially the variant with $\Delta E = 1$ is accepted by about 30% probability and $\Delta E = 2$ variant by about 9%. Horizontal axis represents the second phrase of each sentence. Vertical axis represents the frequency of the acceptance over 1000 generations monitored for a family. Blue, Red and Green bars represent the frequencies for the meanings X, Y and Z, respectively.

The results of the simulation are shown in Fig. 2. We see that the second phrase is forced to be 0 by the selection pressure of the society for large $N$.

Although the language learned from the parent in a family fluctuates, in a large society all children learn almost the same language whose cost is lowest.

## 4c. Communication Error in Society

In 4b the selection pressure of the society which favors the majority is considered. Here we model the communication between pairs of the members in the society. As the smallest information we consider the order of two phrases.

First we summarize our scenario for the communication:
- In a single passage of a sentence we consider the errors in both the sender and the receiver of the sentence.
- Under the situation where both the original and the variant are used, the possibility arises that these two sentences are linked to similar but slightly different meanings. Since we consider the order of two phrases here, one can be used as an affirmative sentence and the other can be used as an interrogative sentence. We interpret such a situation in terms of a grammar.

The communication between pairs is simulated as follows.
(1) We skip the domestic learning in families.
(2) In the first generation all participants start the communication with the same sentence $(1,2)$. The number of the participants is $N$ and we set $N = 100$.
(3) In one generation we consider $N \cdot (N-1)$ times passage. In one passage we choose a pair of the sender and the receiver randomly.
(4) The sender with the sentence $(p_1, p_2)$ sends $(p_1, p_2)$ to the receiver by the probability $1 - P_E$ and $(p_2, p_1)$ by $P_E$ where $P_E$ is the error probability.
(5) The receiver reverses his order of the sentence by the probability $P_R$ if his order is different from the received sentence.
(6) We have repeated the above procedure up to 1000-th generation.

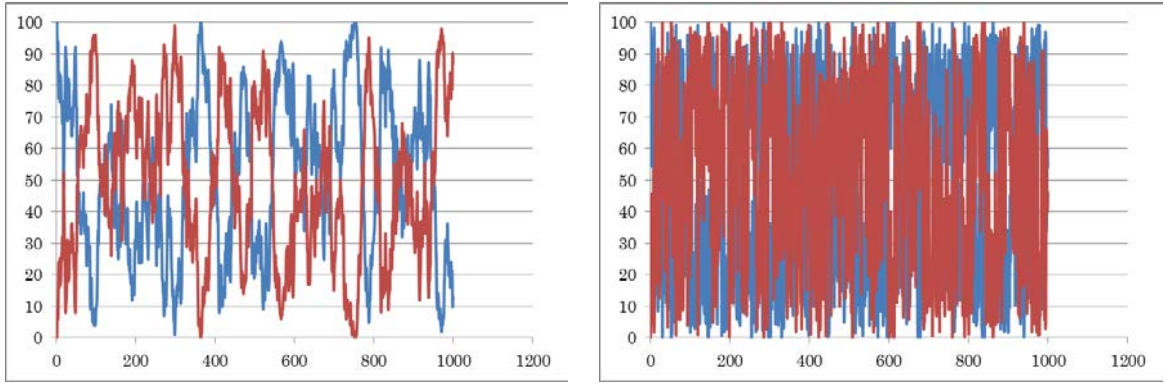

**Fig. 3** Numbers of participants with normal sentence (1,2) in blue and the reversed sentence (2,1) in red for relatively low error probability $P_E$. Horizontal axis represents the generation. Vertical axis represents the number of participants. Left: $(P_E, P_R) = (0.01, 0.01)$ and Right: $(P_E, P_R) = (0.01, 0.3)$.

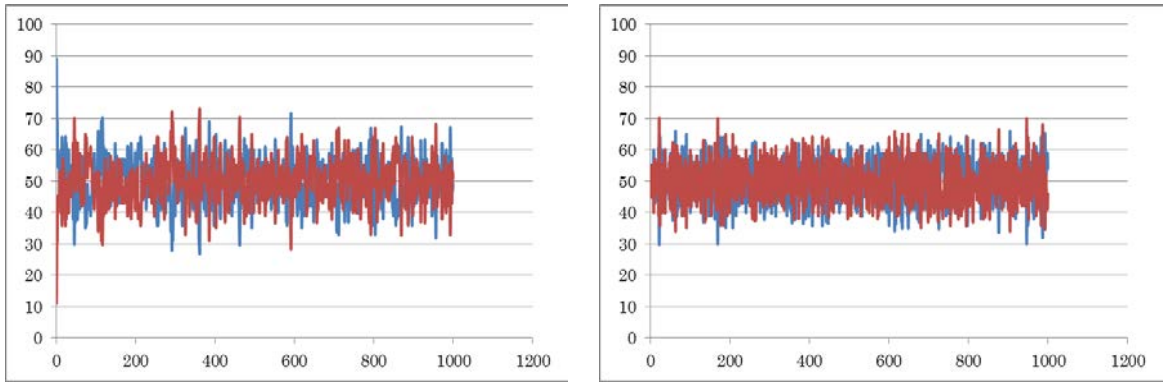

**Fig. 4** Numbers of participants with normal sentence (1,2) in blue and the reversed sentence (2,1) in red for relatively high error probability $P_E$. Horizontal axis represents the generation. Vertical axis represents the number of participants. Left: $(P_E, P_R) = (0.3, 0.01)$ and Right: $(P_E, P_R) = (0.3, 0.3)$.

The results of the simulation are shown in Fig. 3 and Fig. 4. For relatively low error probability $P_E = 0.01$ the order of the phrase fluctuates from generation to generation. On the other hand, for relatively high error probability $P_E = 0.3$ the normal order and the reversed order coexist. Since we do not expect a low error probability for the sender, the coexisting of the normal and reversed orders is a natural consequence.

For our simplest sentences, (1,2) and (2,1), both represent the same meaning. Such a degeneracy allows the coexistence of both. However, if the situation where (1,2) is used as an affirmative sentence and (2,1) as an interrogative sentence is realized, this differentiation breaks the degeneracy and introduces a unique order which is robust against the error.

## 4d. Category Formation via Communication Game

In 4c the change in the order of the phrases in the communication is considered. Here we consider an "apparent" category formation via a communication game.

To establish a grammar an ability to recognize a category is necessary. On the other hand, we do not expect such an ability for agents in our simulations. However, some behavior of the agents is seen as if they utilize categories. Namely, they do not know categories but we see categories there.

First we summarize our scenario for the "apparent" category formation:
- First the vocabulary of the agents is poor so that slightly different states 1 and 2 are represented by one signal 1.
- At one time the sender of the signal 1 expects the action 1 of the receiver. At another time the sender of the signal 1 expects the action 2 of the receiver. The receiver of the signal 1 does either action 1 or 2 without confidence. In this case the demand of the sender is not always met. This situation causes a frustration among the agents.
- If another signal 2 is introduced related to the state 1 or 2, there arises a chance to distinguish these.
- Via a feedback of the communication between the sender and the receiver the usage of the signals 1 and 2 is fixed. One relates the state 1 to the action 1 and the other 2 to 2.
- Since the states 1 and 2 are similar, the signals 1 and 2 are similar. Here we see a category. This "apparent" category is a result of the communication and a trace of the poor vocabulary. The agents do not need category to communicate.

Namely, the correspondence [state$\Rightarrow$signal$\Rightarrow$action] changes by the introduction of a new signal and the accumulated communication with a feedback: in the initial stage [1 or 2$\Rightarrow$1$\Rightarrow$1 or 2], in the intermediate stage [1 or 2$\Rightarrow$1 or 2$\Rightarrow$1 or 2] and the final stage [1$\Rightarrow$1$\Rightarrow$1] and [2$\Rightarrow$2$\Rightarrow$2] separately.

The communication game is simulated as follows. Our scheme is similar to the one employed by Skyrms (Skyrms 2010).

(1) We consider a pair of sender and receiver.
(2) In one process of the communication the state of the sender, 1 or 2, is chosen randomly, he sends a signal to the receiver, and the receiver takes an action.
(3) The sender uses the signal 1 with the probability $r_1$ and 2 with $r_2$ for the state 1.
The sender uses the signal 1 with the probability $r_3$ and 2 with $r_4$ for the state 2.
The receiver takes the action 1 with the probability $r_5$ and 2 with $r_6$ for the signal 1.
The receiver takes the action 1 with the probability $r_7$ and 2 with $r_8$ for the signal 2.
Here $r_1 + r_2 = 1$, $r_3 + r_4 = 1$, $r_5 + r_6 = 1$ and $r_7 + r_8 = 1$.

(4) Introducing the score variable $x_i$ ($i=1,3,5,7$) the probability $r_i$ ($i=1,3,5,7$) is determined by $r_i = \frac{1}{2}\left[1 + \tanh\frac{x_i}{2}\right]$. Before starting the game we set $x_1 = x_3 = x_5 = 0.42$ to realize $r_1 = r_3 = r_5 \approx 0.7$ and set $x_7 = 0$ to realize $r_7 = 0.5$. The error in the communication is represented by the nonzero value of $r_2$, $r_4$ and $r_6$. Initially unknown signal 2 leads to a tentative action, 1 or 2, with equal probabilities, $r_7 = r_8 = 0.5$.

(5) At the end of the one communication process the score $x_i$ increases or decreases reflecting the effectiveness of the communication. We have assumed that our agents have an ability of a feedback and can improve the communication. The process $[1 \Rightarrow 1 \Rightarrow 1]$ is fully effective so that $x_1$ and $x_5$ increase by 0.64. The process $[1 \Rightarrow 1 \Rightarrow 2]$ is not effective so that the scores are unchanged. The process $[1 \Rightarrow 2 \Rightarrow 1]$ is partially effective so that $x_7$ increases by 0.16 but $x_1$ decreases by 0.16. The process $[1 \Rightarrow 2 \Rightarrow 2]$ is not effective so that the scores are unchanged. The success of $[2 \Rightarrow 2 \Rightarrow 2]$ leads the increase of $r_4$ and $r_8$ so that $x_3$ and $x_7$ decrease by 0.32. The process $[2 \Rightarrow 2 \Rightarrow 1]$ is less successful so that $x_3$ decrease by 0.16 but $x_7$ increases by 0.16. The unsuccessful process $[2 \Rightarrow 1 \Rightarrow 1]$ leads to the increases of $x_3$ and $x_5$ by 0.08. The less unsuccessful process $[2 \Rightarrow 1 \Rightarrow 2]$ leads to the increases of $x_3$ by 0.24 and the decrease of $x_5$ by 0.24.

(6) We have repeated the above communication process 100 times.

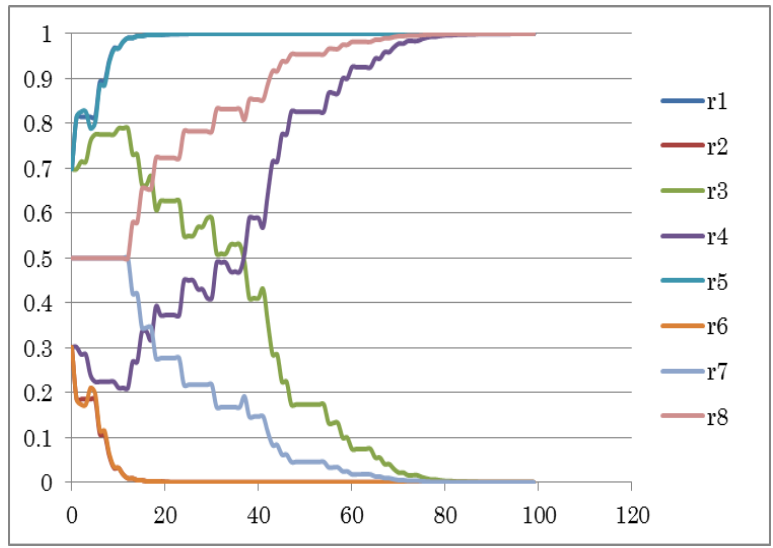

**Fig. 5** Probabilities for selecting signals ($r_1, r_2, r_3, r_4$) and actions ($r_5, r_6, r_7, r_8$). Horizontal axis represents the number of experienced communication. Vertical axis represents the probability.

The results of the simulation are shown in Fig. 5. The behaviors of the sender and the receiver are not error free but converge to a new pattern where the introduction of the new signal 2 enables to distinguish the states 1 and 2. The convergence is brought by the repeated feedback under the bias.

The states 1 and 2 are similar and initially represented by one signal 1. The development here is that these two states become to be represented by different signals. The agents do not group the states to form a category, though we expect such an ability of grouping for the category formation. On the other hand, the initial vocabulary of our agents is too poor to distinguish the states so that a group of states is represented by a common signal. They do not form a group but differentiate the states in it. Thus the development here is a reverse process of the expected category formation.

## 5. Summary

We have demonstrated the following using a series of model simulations:
- a. Diversity generated by error
- b. Stabilization by majority formation in society
- c. Background of grammar by shuffling of order
- d. Differentiation of not intentional category.

We have assumed that a primitive language is not logically constructed by a grammar but a product of the interaction between individual and society. The physical condition of the individual is an important factor for the language. If there exist a mechanism to amplify the success of the communication, a robust language in the society is established easily.